# Fully quantum mechanical dynamic analysis of single-photon transport in a single-mode waveguide coupled to a traveling-wave resonator

Edwin E. Hach III[1], Ali W. Elshaari[2], Stefan F. Preble[2]

[1]Department of Physics, Rochester Institute of Technology, 85 Lomb Memorial Drive, Rochester, New York 14623, USA

[2]Microsystems Engineering, Rochester Institute of Technology, 77 Lomb Memorial Drive, Rochester, New York 14623, USA

## Abstract

We analyze the dynamics of single photon transport in a single-mode waveguide coupled to a micro-optical resonator using a fully quantum mechanical model. We examine the propagation of a single-photon Gaussian packet through the system under various coupling conditions. We review the theory of single photon transport phenomena as applied to the system and we develop a discussion on the numerical technique we used to solve for dynamical behavior of the quantized field. To demonstrate our method and to establish robust single photon results, we study the process of adiabatically lowering or raising the energy of a single photon trapped in an optical resonator under active tuning of the resonator. We show that our fully quantum mechanical approach reproduces the semi-classical result in the appropriate limit and that the adiabatic invariant has the same form in each case. Finally, we explore the trapping of a single photon in a system of dynamically tuned, coupled optical cavities.

## I. INTRODUCTION

Light speed transmission and low noise properties make photons indispensable for quantum communication and information processing. In free space optics, a quantum bit (or qubit) of information can be manipulated and encoded in any of several degrees of freedom, notably polarization, in which case this process is usually straightforward using birefringent waveplates [1].

In order to have more functionality in future quantum computing systems, devices need to be scaled down to the micro- and nano- integration level. One potential platform is Silicon, which has desirable optical properties for integrated optical systems at the telecommunication wavelength of 1550 nm. In addition, it is considered as a candidate for generating single photon sources relying on the high third order nonlinearity $\chi^{(3)}$ [2]. Using such sub-Poissonian sources enables revolutionary new technologies [3]- individual photons have been used to dramatically enhance communication security [4], have increased measurement precision beyond the standard quantum limit [5,6], have been used to beat the diffraction limit[7,8], and they hold great promise for quantum computation [9,10]. Surprisingly a quantum mechanical theory describing photon/resonator interactions solved under the steady-state harmonic excitation condition has only recently been formulated [11,12]. Further, the full dynamical behavior of the system needs to be considered in order to describe more complex single photon manipulation processes. In this work we develop a dynamical model for single photon interactions with cavities. As a first application to this approach, we describe the process of single photon energy lifting in optical cavities, the semi-classical analogue of which has already demonstrated [13,14,15]. We further show that fully quantized model of the process follows the adiabatic condition for dynamical systems and that the process has 100% wavelength conversion efficiency for states trapped in the cavity during the tuning process. As a second application for our solutions we present and analyze single photon trapping analogue to coherent population trapping (CPT) using tunable micro-cavities which presents the building blocks of optical memories on chip [16,17,18,19]. The adopted design employs a lossless storage unit approach with a tuning mechanism compatible with common silicon photonic circuits.

This paper is organized as follows. In Sec. II we describe the interaction of single photons with traveling-wave optical cavities. We start by describing the quantum electrodynamic effective Hamiltonian for the system and then reviewing the stationary state solutions. Next we present a full dynamical model of the interaction between an arbitrary, propagating, single photon “wave packet” and the cavity-waveguide system. Throughout, for mathematical simplicity, we adopt a single photon state with an initial Gaussian spatial amplitude distribution as our representative photon “wave packet.” We remark here that the use of quotation marks is to remind the reader that this is function is not a position representation “wave function” for the photon, as no such first-quantized representation is possible for the quantized electromagnetic field. In Sec. III, we analyze the energy lifting property of photons in active optical cavities. In Section IV, we show that single photon can be trapped efficiently in a system of dynamically tuned optical cavities. In Section V, we present a summary of the work and a brief discussion of our future plans for studying photonic circuits. Finally, in the Appendix, we show our analytical method for deriving the steady state transport solutions in an "ansatz-free" manner.

## II. SINGLE PHOTON- CAVITY DYNAMICS

### *a. Theoretical Model*

Consider a one-dimensional, single mode waveguide coupled to cavity via an evanescent coupling, as shown in Fig. 1. In what follows, we assume that the single photon state is injected into the waveguide from the left and propagates through the waveguide (cavity) in the positive $x$ (counter-clockwise) direction. Specifically, we assume that there is no impurity interaction within the system and therefore no contribution due to reflection, as easily verified by "turning off" the impurities in Ref. [12] by setting $g_a = g_b = h = 0$ in the results presented in that paper. Setting $\hbar = 1$, as we will do throughout unless otherwise stated, the effective Hamiltonian for the system we study takes the form [12]

$$\hat{H}_{\text{eff}} = \int dx c^{+}(x)\left(\omega_0 - i v_g \frac{\partial}{\partial x}\right)c(x) + \left(\omega_{\text{c}} - i\frac{1}{\tau_{\text{c}}}\right)\hat{a}^{+}\hat{a} + \int dx \delta(x)\left[V c^{+}(x)\hat{a} + V^{*}\hat{a}^{+}c(x)\right] \tag{1}$$

where $\left(\hat{c}(x), \hat{c}^{+}(x)\right)$ are the position dependent inverse Fourier representations of the usual Boson ladder operators describing the rightward traveling waveguide mode, and $\left(\hat{a}, \hat{a}^{+}\right)$ are the Boson ladder operators describing the counter-clockwise cavity mode. The canonical commutation relations for the system are $\left[\hat{a}, \hat{a}^{+}\right] = 1$, $\left[\hat{c}(x), \hat{c}^{+}(x')\right] = \delta(x - x')$; $\left[a, \hat{c}^{+}(\hat{x})\right]$ and all other combinations vanish. In using this form of the effective Hamiltonian we are assuming that the waveguide is driven at a frequency within a narrow range far from the cutoff frequency of its dispersion relation, and that $v_g$ is the group velocity for the traveling waveguide mode [11]. The evanescent coupling is represented by the local interaction term having coupling strength $|V|$ where the coupler is situated at $x = 0$. The cavity lifetime of the ring-resonator is $\tau_{\text{c}}$ where cavity losses have been included using the simple model of a complex frequency (energy) shift from the resonance frequency, $\omega_{\text{c}}$, of the ring resonator [20]. This simple model for dissipation results in the explicitly non-Hermitian form of the effective Hamiltonian.

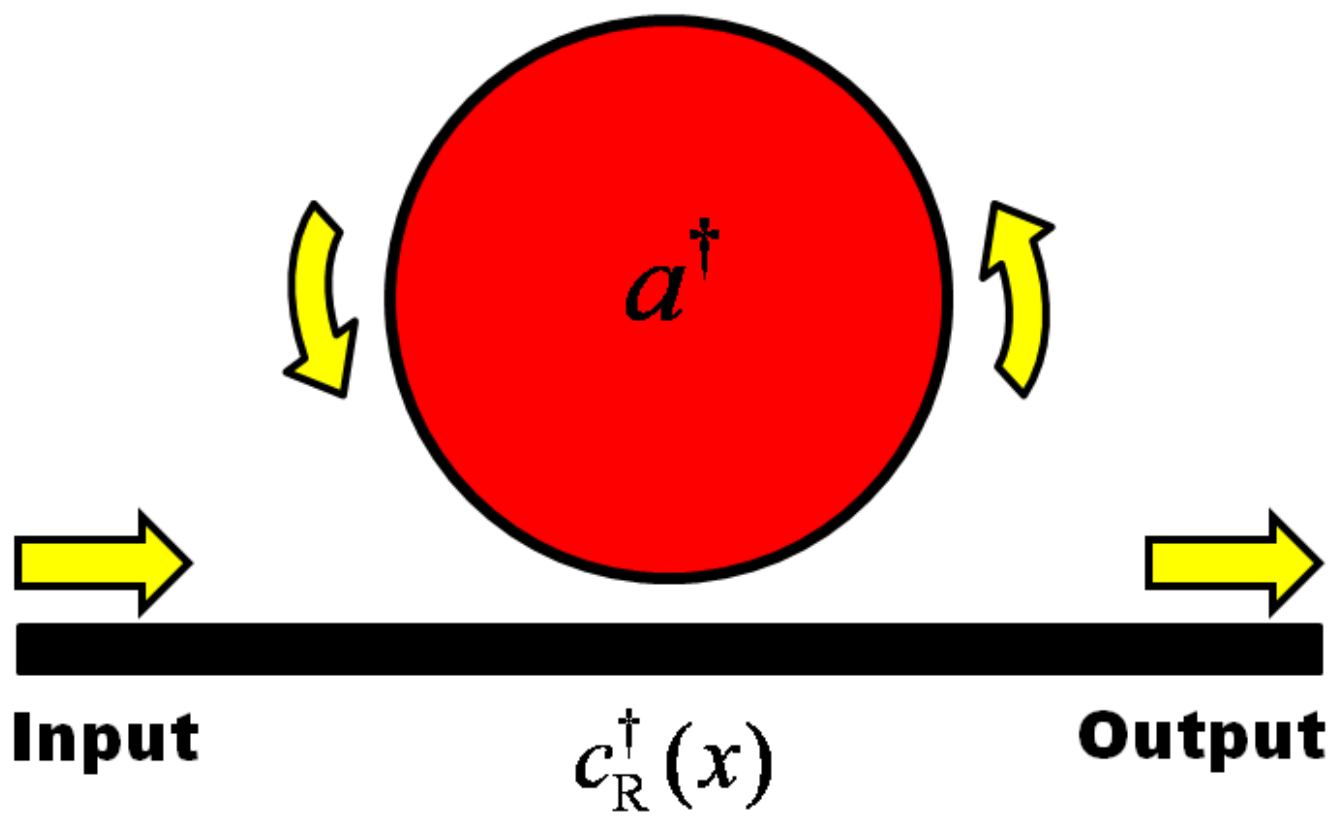

Fig. 1. Schematic of waveguide-cavity coupled system.

A general one photon state of the system can be written in the form

$$|\Phi_1(t)\rangle = \left(\int dx\tilde{\phi}(x,t)\hat{c}^+(x) + \tilde{e}_{\text{cav}}(t)\hat{a}^+\right)|0,0\rangle \tag{2}$$

where $\tilde{\phi}(x,t)$ and $\tilde{e}_{\text{cav}}(t)$ are the time dependent excitation amplitudes for field in the waveguide and the cavity, respectively, and where $|0,0\rangle \equiv |0\rangle_{\text{w.g.}} \otimes |0\rangle_{\text{cav}}$ represents the vacuum state of the field.

The quantum dynamics of the system is described by the Schrödinger Equation

$$i\frac{\partial}{\partial t}|\Phi_1(t)\rangle = \hat{H}_{\text{eff}}|\Phi_1(t)\rangle \tag{3}$$

Substitution of the second quantized forms in Eqns. (1) and (2) into Eqn. (3) and then projecting alternatively on to the one photon waveguide ("w.g.") and cavity ("cav") subspaces using the single photon basis states $|1,0\rangle = \hat{c}^+(x)|0,0\rangle$ and $|0,1\rangle = \hat{a}^+|0,0\rangle$, respectively, yields the coupled set of time evolution equations for the first quantized excitation amplitudes,

$$\left(\omega_0 - iv_g\frac{\partial}{\partial x} - i\frac{\partial}{\partial t}\right)\tilde{\phi}(x,t) + \delta(x)V\tilde{e}(t) = 0 \tag{4}$$

$$\left(\omega_c - i\frac{1}{\tau_c} - i\frac{\partial}{\partial t}\right)\tilde{e}_{\text{cav}}(t) + V^*\tilde{\phi}(0,t) = 0 \tag{5}$$

These are precisely the dynamical equations derived by Shen and Fan in Ref. [12] suitably modified for the system we study here, and, in fact, our choice of notation is intentionally similar to that adopted by Shen and Fan so that the reader can more easily translate between our results and the existing literature in this area.

### *b. Stationary state solutions*

For ease of reference we now include a brief review of the stationary state analysis of the system we consider. One seeks stationary states of a quantum dynamical system by seeking solutions to Eqn. (3) having the form $\left|\Phi_1(t)\right\rangle = \left|\phi\right\rangle e^{-i\omega t}$ where $\omega$ is the eigenfrequency for the system, related the energy eigenvalue, $\varepsilon$ , in the usual way, $\varepsilon = \hbar\omega$. Doing this results in the coupled set of equations

$$\left(\omega_0 - iv_g \frac{\partial}{\partial x} - \omega\right)\phi(x) + \delta(x) V e_{\rm cav} = 0 \qquad (6)$$

$$\left(\omega_{\rm c} - i\frac{1}{\tau_{\rm c}} - \omega\right)e_{\rm cav} + V^* \phi(0) = 0 \qquad (7)$$

where the time independent amplitudes are defined via $\tilde{\phi}(x,t) = \phi(x)e^{-i\omega t}$ and $\tilde{e}_{\rm cav}(t) = e_{\rm cav}e^{-i\omega t}$ . Typically, the stationary state solution for the waveguide excitation amplitude is written in the form $\phi(x) = e^{iQx}\left[\theta(-x) + t\,\theta(x)\right]$, where $Q$ is the wave vector for the traveling mode, $t$ is the transmission coefficient for the traveling mode after interacting with the ring-resonator, and $\theta(x)$ is the Heaviside step function. This form follows from the single particle Bethe Ansatz for the interacting eigenstate once the Lipmann-Schwinger formalism is used to identify the input and output amplitudes for the waveguide state vector [21]. In this case the Bethe Ansatz is especially simple as the waveguide state is taken to be a momentum eigenstate $(p = \hbar Q)$, an approximation imposed in order to restrict the waveguide to a single, dominant mode. In the Appendix we present the details of an alternate and, we believe, equally direct and slightly more general method for finding the stationary state solutions for the system and the associated momentum eigenvalue (scaled by $\hbar$ ) and transmission amplitude. We quote the results here for use in selecting operating parameters for our dynamical simulations below:

$$Q = \frac{\omega - \omega_0}{v_g} \qquad (8)$$

$$t=\frac{\omega-\omega_{c}+i\frac{1}{\tau_{c}}-i\Gamma}{\omega-\omega_{c}+i\frac{1}{\tau_{c}}+i\Gamma} \tag{9}$$

where $\Gamma\equiv\frac{|V|^{2}}{2v_{g}}$.

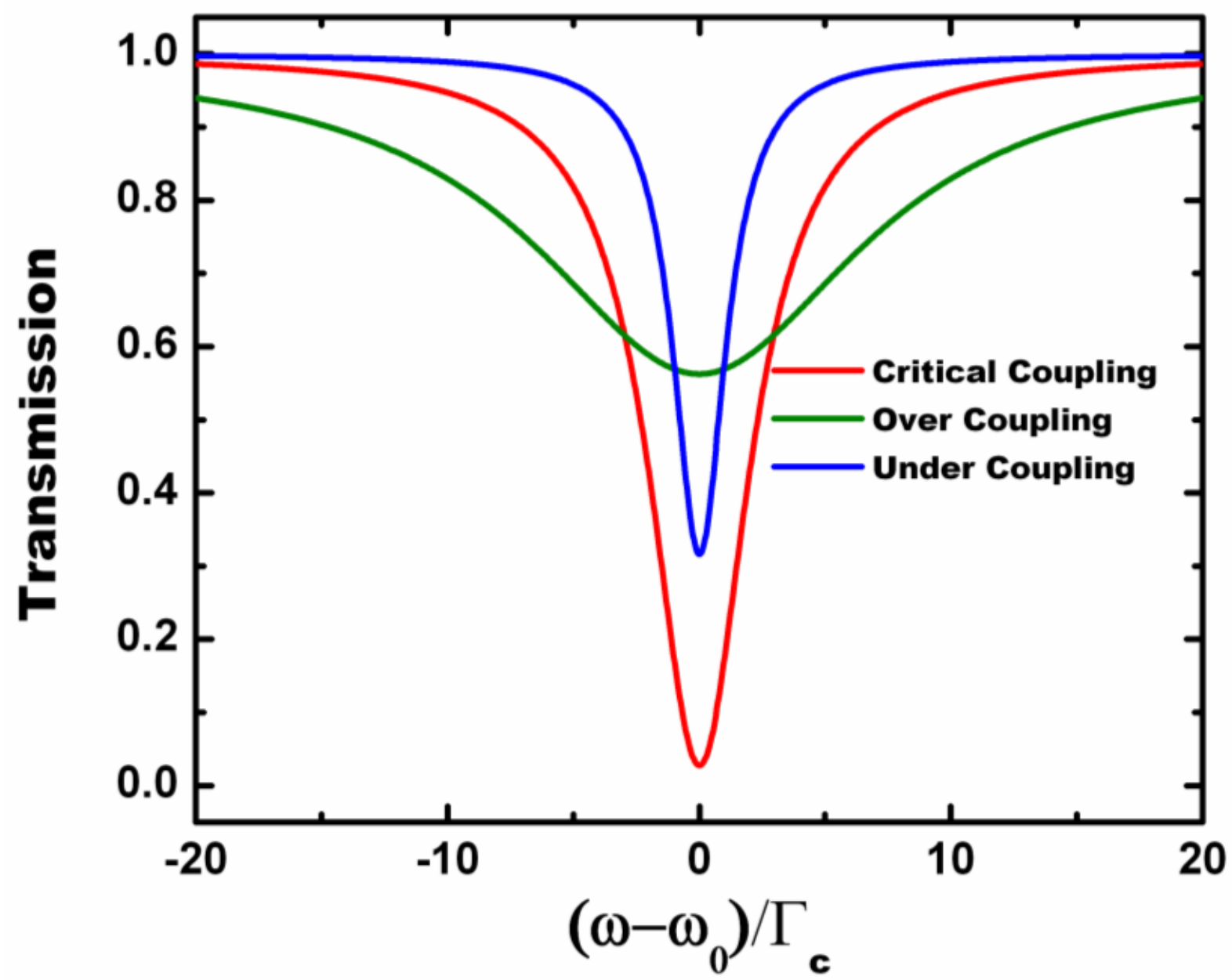

FIG. 2. Transmission of the cavity under different coupling conditions.

We now choose determine model parameters that expose the salient features of the time evolution of the single photon state. The transmission of the waveguide cavity systems depends on the coupling condition with respect to the internal cavity loss. Fig. 2 shows the single photon state transmission for critically coupled ( $\Gamma=\frac{1}{\tau_{c}}$ ), under-coupled ( $\Gamma<\frac{1}{\tau_{c}}$ ), and over-coupled ( $\Gamma>\frac{1}{\tau_{c}}$ ) cavity waveguide system. For demonstration purposes the cavity is operated slightly in the under-coupled regime to provide high enough photon life times for different dynamic processes. In practical devices, complete capture of the wave-packets can be achieved [22].

### *c. Numerical analysis of dynamic interactions*

In order to examine the transient dynamic response of the system to an arbitrary single photon input, we solve the equations of motion (Eqn. (4 and 5)) numerically. To do so we use a finite difference method, which is commonly used to analyze electromagnetic scattering and propagation [23]. We begin by specifying an initial input single-photon state. The time evolution of the state is described by using the knowledge of the amplitude functions at each grid point in space at the previous time step and propagating forward to the next time step using the finite difference approximation shown in Eqns. (10 and11).

$$\tilde{\phi}(k, n+1) = \Delta t \cdot \left(\frac{\tilde{\phi}(k,n)}{\Delta t} - i\omega_0 \tilde{\phi}(k,n) - \left(\frac{\tilde{\phi}(k,n) - \tilde{\phi}(k+1,n)}{\Delta x / v_g}\right) - i\delta(N) V \tilde{e}(n)\right) \tag{10}$$

$$\tilde{e}(n+1) = \Delta t \cdot \left(\frac{\tilde{e}(n)}{\Delta t} - i\omega_c \tilde{e}(n) - iV\tilde{\phi}(N,n) - \frac{\tilde{e}(n)}{\tau_c}\right) \tag{11}$$

Where N is the spatial grid point representing the cavity/waveguide coupling region; the interaction is clearly a local one.

To demonstrate this process and to elucidate the quantum mechanical features of the system we consider a single photon state with a Gaussian amplitude (hereafter referred to as the "Gaussian wave packet") incident on the cavity waveguide system. The propagation of the single photon wave packet through the system is typified by the profiles plotted in Fig. 3.

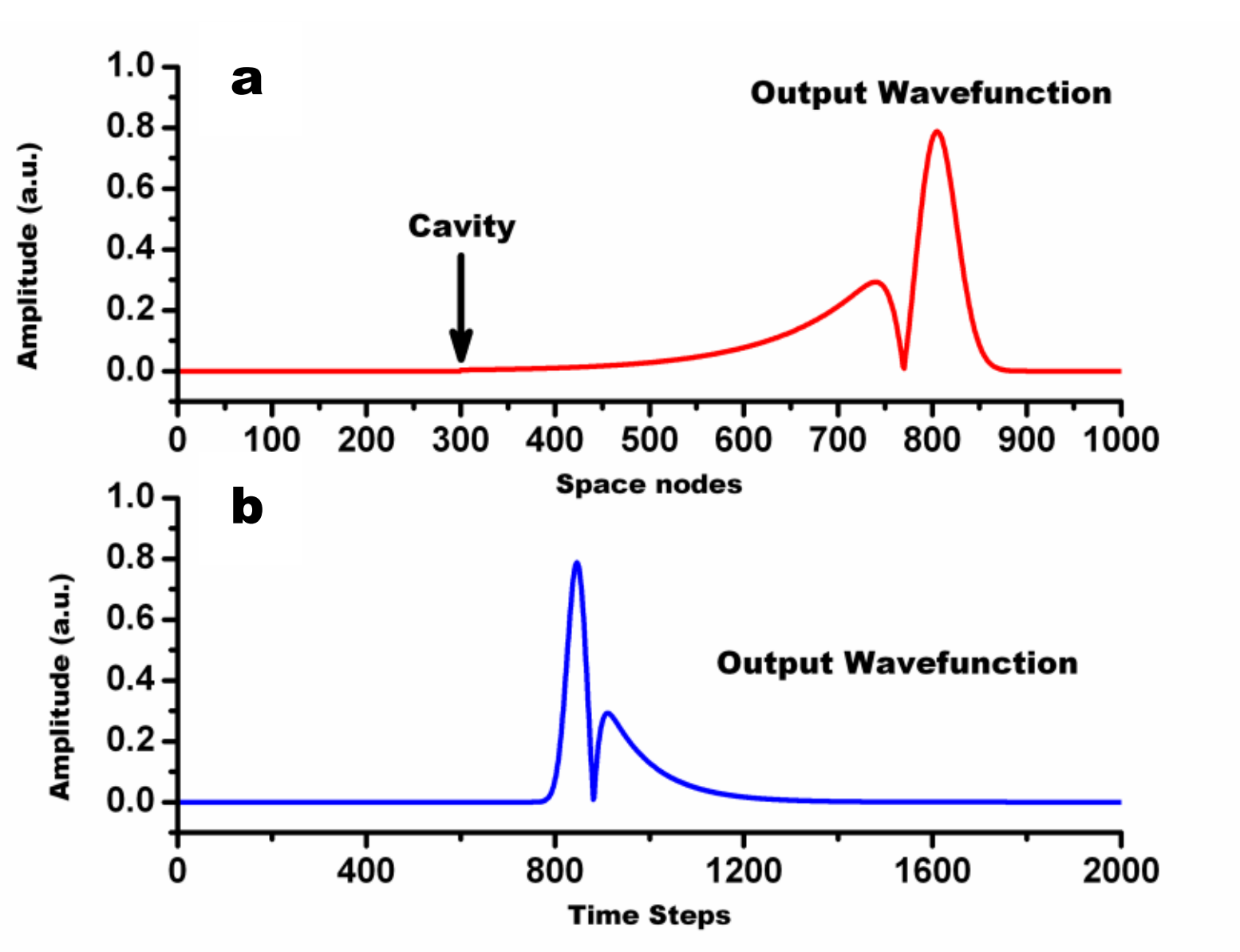

FIG. 3. Transmission of a Gaussian packet through the cavity waveguide system.

In Figure 3a we display the absolute magnitude of the probability amplitude for the wavepacket as a function of position along the waveguide for a fixed time. We have chosen that time to be long after the photon could have (i) first interacted with the microcavity resonator at the coupling region and/or (ii) could have completed a round trip in the resonator if it were injected upon interaction at the coupler. Several general features of the solution are apparent. First, there are clearly general positions within the waveguide at this time at which the photon is likely to be detected. Unlike the case involving classical fields, this is a signature of quantum interference between the branch of the photon *state vector* representing direct transmission of the photon at the coupler and the branch representing a single round trip in the cavity before transmission. Put another way, if we were to place perfectly efficient single photon detectors at the locations of the two peaks in the figure, then (at most, see below) only one of them would fire in any individual run of the experiment. By operating in the under-coupled regime, we can neglect the contributions of higher order terms arising form more than one round trip through the microcavity prior to transmission through the waveguide. Second, it is the larger peak in Figure 3a that corresponds to direct transmission. This peak is centered at $x_d = v_g t$ where $t$ is the time at which the "snapshot" in Figure 3a is taken and we are taking $t = 0$ to be the moment at which the photon is incident upon the coupling region. The round trip peak is smaller, broader and lags the direct peak by a distance of $\Delta x = v_g(2\pi nR/c)$ where $R$ is the radius of the microcavity and $n$ is the effective intra-cavity index of refraction. Third, the round trip peak is attenuated relative to the direct peak; this is a result of (i) our choice of coupling strength (which is unitary and therefore probability conservative) and (ii) cavity losses (which is irreversible and therefore not probability conservative). The unitary source of the difference in peak height is a simple consequence of our choice to operate in the under-coupled regime. The irreversible part is due to cavity losses. This means that in any individual realization of the experiment there is a finite probability for the photon to be "lost" to the environment. Because we are considering only single roundtrip events in our model with relatively weak losses, we expect that the resultant lack of normalization in the output will be small. Because it does not impact the major results presented in this paper, we make no effort to deal with this loss quantitatively here, but it is readily apparent by inspection of Figure 3a that it is a small effect in the regime we are considering. To examine a fourth feature of the solution, consider Figure 3b, which shows the time evolution of the modulus of the amplitude function at a fixed position in the waveguide beyond the coupling region. Clearly, the direct peak arrives at an earlier time that the roundtrip peak, as discussed above. One can think of this curve in terms of a photon counting Gedankenexperiment. Suppose we measure the probability of detecting $m$ photons at a time delay $T_d$ after the interaction, $P(m,T_d)$ for an ensemble of similarly prepared runs of the single photon transport experiment. Figure 3b suggests we should see non-zero results only around $P(1,T_d \approx x/v_g)$ and $P(1,T_d) \gtrsim x/v_g + 2\pi nR/c$. Again, these results are ultimately traceable to quantum interference between the branches of the single photon state vector. Further, the decay tail of the second peak depends on the cavity decay rate and the coupling factors between the cavity and the waveguide.

We envision that through careful quantitative study of the features described in this section, we can, with the advent of efficient, single photon detectors, develop an experimental protocol for characterizing the optical properties of the photonic structures we are considering here. This is an exciting possibility that we will explore elsewhere. Instead, for the purpose of this paper, we now apply our dynamical single solution to the study the operation of a couple of photonic devices of interest.

## III. ADIABATIC MODULATION OF SINGLE-PHOTON PACKETS

### *a. Theoretical Model*

In this section we examine the possibility of dynamically tuning the wavelength of a *single photon* via adiabatic following. The model that we are considering here is similar to the one in section II except that we now allow for the dynamic tuning of the resonance frequency of the ring resonator, mathematically by allowing $\omega_c$ in the effective Hamiltonian (Eqn. (1)) to become a parametric function of time, $\omega_c(t)$. The system is shown schematically in Fig. 4 where the change in the color of the cavity from its color at $t_0$ depicts the change of the cavity resonance frequency. Clearly, the parametric change in the system affects only the harmonic oscillator (free ring resonator) term of the effective Hamiltonian. We expect, then, that for a sufficiently slowly varying function $\omega_c(t)$, the wavelength of the cavity will adiabatically follow the parametric shift in the resonance frequency. This has been demonstrated theoretically and experimentally, by, among others, one of the present authors (SP), in the case of classical electromagnetic waves [13,14,15]. The effect is a clear manifestation of the well-known adiabatic theorem [20], and our mission here is to show, via direct computation, that the phenomenon carries over directly to the fully quantum mechanical case of single-photon transport.

It is interesting to note that for the harmonic oscillator, the fully quantum mechanical import of the adiabatic theorem is fully accessible from semi-classical analysis. Ultimately this is related to the fact that the harmonic oscillator is one of the systems that has perfectly closed orbits in phase space, the trajectory being the energy ellipse, and the classical adiabatic invariant being the action integral, $\oint pdq = \frac{U}{\omega}$, where $U$ represents the total mechanical energy of an oscillator having natural angular frequency, $\omega$, and the action integral is geometrically equivalent to the area of phase space bounded by the energy ellipse. Semi-classically, the action is related to this adiabatic invariant via the Bohr-Sommerfeld quantization condition from the "old quantum theory," $\oint pdq = N\hbar$. Combining the two expressions for the action integral clearly results in the simple energy quantization condition for the quantized oscillator. It is especially important realize in interpreting our single photon results that, although such simple analysis cannot tell us about the actual quantum *state* of the photon field, in the case of the harmonic oscillator (a realization of which formally represents the single-mode photon field) it does produce exactly the energy *eigenvalues*. It is for this reason that the adiabatic invariant for the fully quantum mechanical Hamiltonian is exactly the same as for the case involving classical fields; this is in fact the underlying reason why the authors of Ref. [24] were able to explain exactly the adiabatic energy shift of a *photon* using only *semi-classical* analysis. Using simple Δ-calculus, it is clear that for any simple harmonic oscillator $0 = \frac{\Delta\left(U/\omega\right)}{U/\omega} = \frac{\Delta U}{U} + \frac{\Delta\omega}{\omega} = \frac{\Delta U}{U} - \frac{\Delta\lambda}{\lambda}$, where the second equality results from our assumption that the oscillator represents a cavity photon, $\omega = \frac{2\pi c}{\lambda}$. Clearly we must have, semi-classically and quantum mechanically, $\frac{\Delta\lambda}{\lambda} = \frac{\Delta U}{U}$. This relationship describes a detailed balance between the energy of the cavity photon and the work done on the system by the external agent that performs the tuning of the cavity resonance; quite aptly, there is a thermodynamically adiabatic transfer of energy from the environment to the cavity.

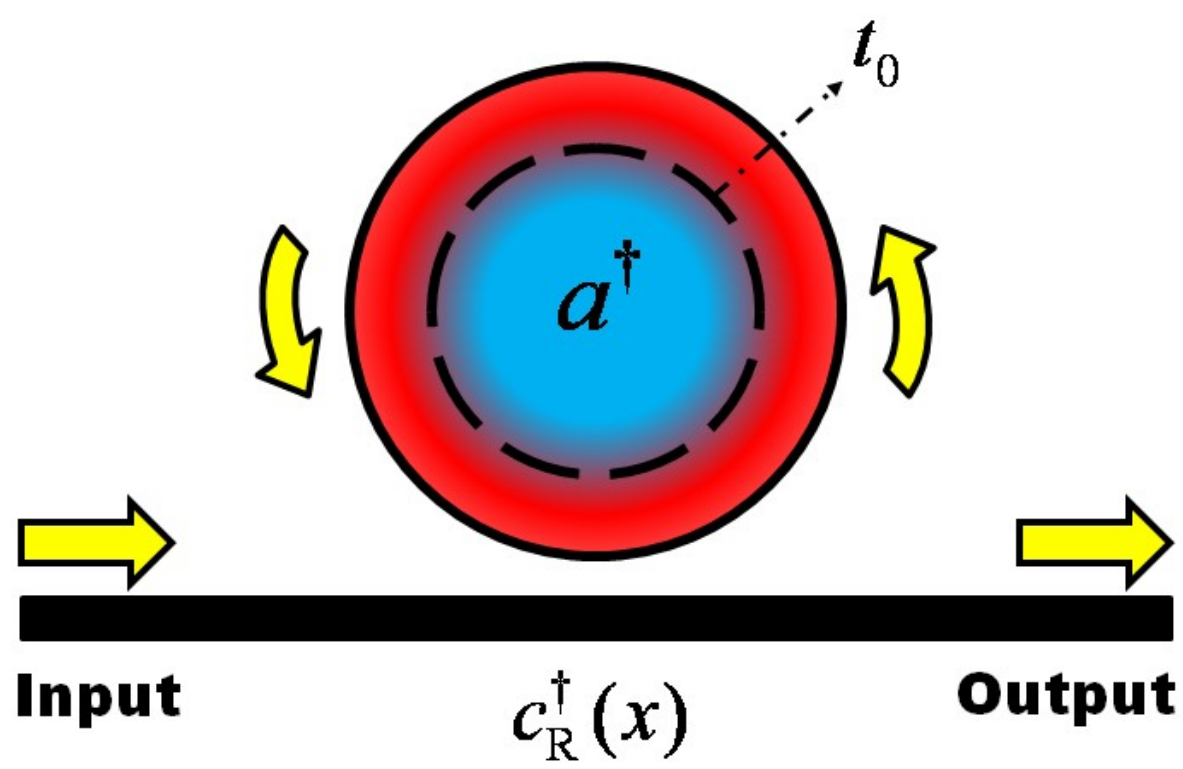

FIG. 4. Schematic of the single photon energy lifter. The optical cavity is adiabatically tuned at time $t_0$ from one stationary state to another.

### *b. Results*

We now directly demonstrate the single photon adiabatic wavelength shift and examine its dynamical behavior. To begin, we repeat Eqns. (4) and (5) with the modification to the cavity resonance frequency $\omega_c$,

$$\left(\omega_0 - iv_g\frac{\partial}{\partial x} - i\frac{\partial}{\partial t}\right)\tilde{\phi}(x,t) + \delta(x)V\tilde{e}(t) = 0 \tag{12}$$

$$\left(\omega_c(t) - i\frac{1}{\tau_c} - i\frac{\partial}{\partial t}\right)\tilde{e}_{cav}(t) + V^*\tilde{\phi}(0,t) = 0 \tag{13}$$

In solving these equations using the numerical algorithm outlined above we will assume a simple linear shift to the resonance frequency, $\omega_c(t) \sim t$. The physical result is insensitive to the exact parameterization, as long as the change is "slow" in comparison with the intrinsic time scale of the system, which in this case set by the inverse of the mode frequency spacing of the ring resonator. Further, it should be experimentally straightforward to dynamically tune a ring resonator resonance at a constant rate over the interesting region of parameter space.

The time evolution equations are solved using the numerical technique described earlier. We first consider wavelength shifts that occur over a time scale smaller than the cavity photon lifetime. In Fig. 5 we show our solutions for the cavity photon amplitude function, $\tilde{e}_{cav}(t)$, for several different values of the adiabatic shift. The evidence of quantum control of the cavity photon wavelength is clear even at the single photon level. We stress that this mechanism for wavelength conversion of a single photon involves only linear optical processes and therefore obviates the need for large nonlinear susceptibilities that is a technological barriers to the development of quantum circuit elements.

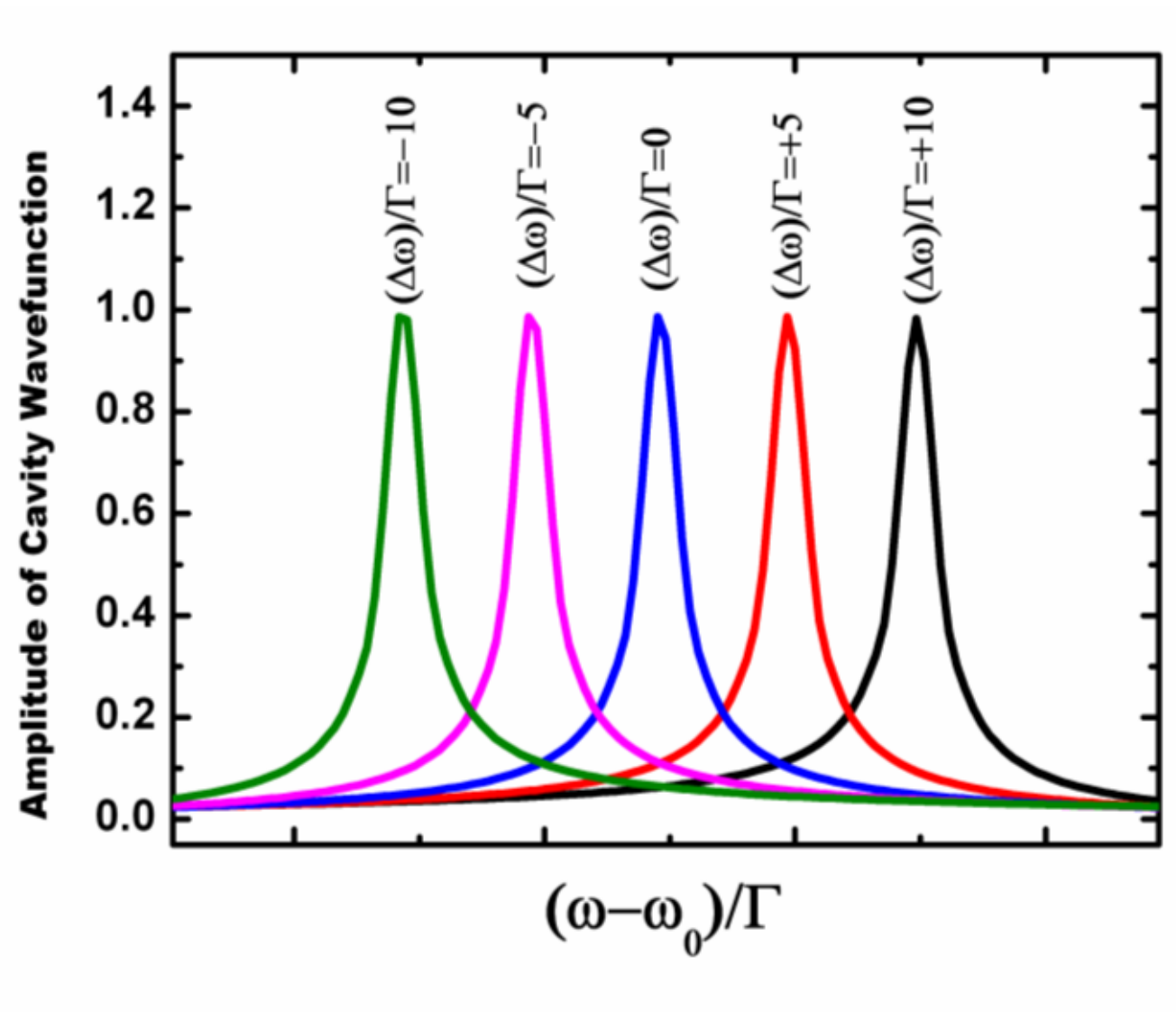

FIG. 5. Single photon energy state changes adiabatically as a function of the cavity tuning frequency.

Next we study the conversion efficiency as a function of the switching time. The results are shown in Fig. 6. The switching function is varied linearly with time from the initial state to the final state. We clearly see that as the switching time approaches the time the photon spends inside the cavity there is a lower probability to detect the photon at the new Eigen state. This probability has an upper limit of 1 when the conversion is done in a time $T << \tau_{c.}$ We can understand the degradation in amplitude as the result of a competition between the adiabatic tuning process and the non-adiabatic effect of the finite cavity decay rate we are considering in our model. If the switching time is short, the adiabatic effect dominates (which may seem counterintuitive, but one must recall that adiabatic does not mean "slow" it means without irreversible energy exchange – it just happens that many text book examples of the adiabatic theorem involve slow processes, [25]). On the other hand, if the switching time is comparable to the cavity lifetime, $T \sim \tau_c$, the effects of cavity decay become apparent. Clearly, if $T << \tau_c$, we face the trivial situation of having a very low probability of there being a photon whose wavelength we can hope to shift. We stress here that we are considering micro-cavities with mode frequency spacing that are much larger than the inverse of the relevant time scale for the adiabatic change, $1/T$. We have yet to apply our single photon model to the case in which other modes are accessible and thus excited when $1/T \sim \Delta\omega$. We expect that, as in the semi-classical case, the wavelength conversion will lose fidelity as a result of the distribution of the injected (or extracted) energy over several modes (as experimentally demonstrated [26]) , in turn leading to an increase in the effective entropy of the system (as more microstates, viz. the other modes become available), thus introducing a non-adiabatic component to the energy transfer. We shall study the details of this more complicated case elsewhere.

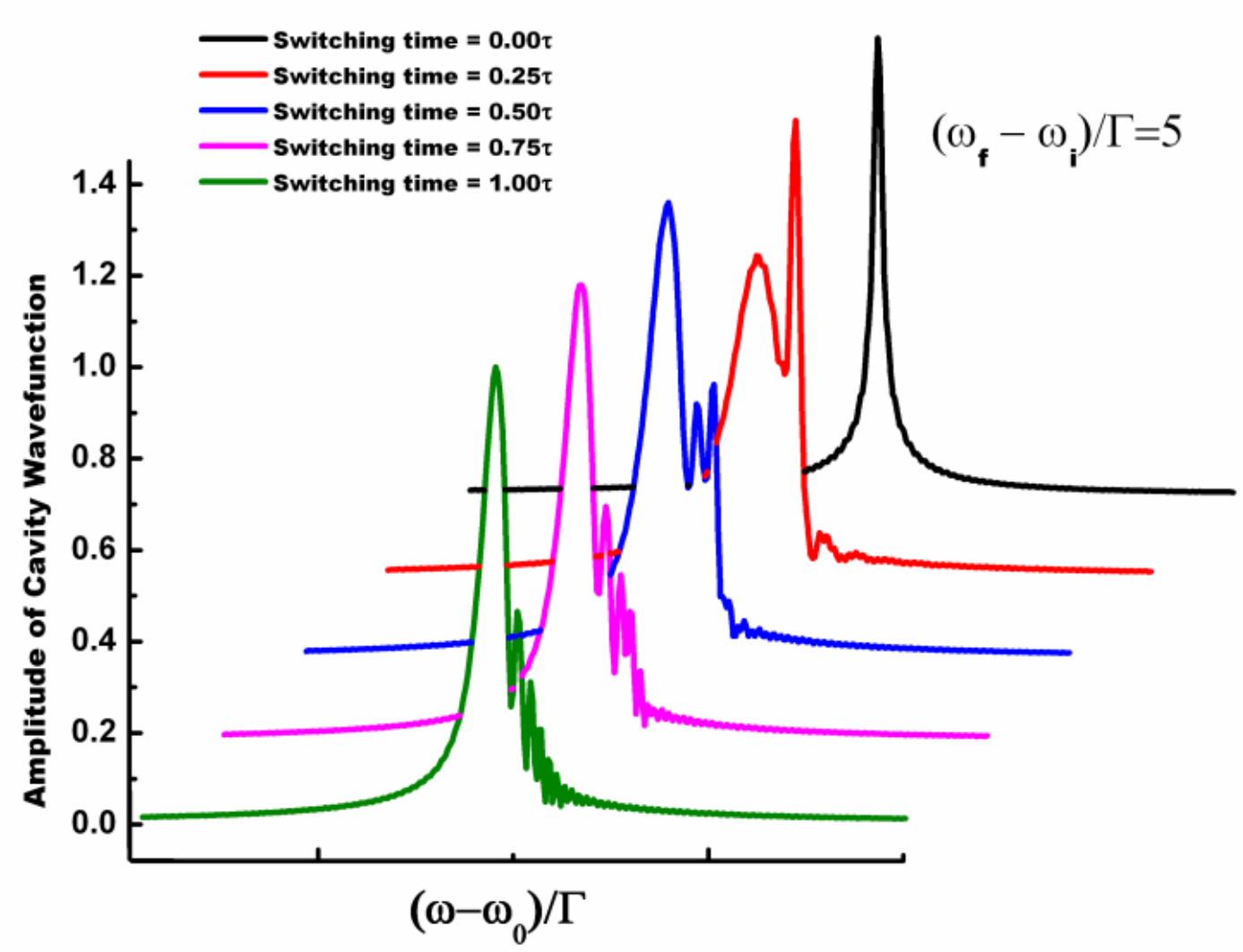

FIG. 6. Shows the final single photon state as a function of the tuning time. The conversion efficiency degrades with slower tuning.

We now analyze another test to verify that the probability of detecting a photon in the new (old) eigenstate after the conversion is $P$=1 ($P$=0). A cavity is placed in series with the dynamic cavity and the experiment is performed twice. Initially, the output cavity is tuned to the original eigenstate and second to the new eigenstate in such a way as to probe the photon energy at the output of the dynamical system. The equations of motion describing the systems can be derived in a way analogous to that discussed in Section II resulting in the system,

$$\left(\omega_0 - iv_g \frac{\partial}{\partial x} - i\frac{\partial}{\partial t}\right)\tilde{\phi}(x,t) + \delta(x_a)V\tilde{e}_a(t) + \delta(x_b)V\tilde{e}_b(t) = 0 \tag{14}$$

$$\left(\omega_{\mathrm{c,a}}(t) - i\frac{1}{\tau_{\mathrm{c,a}}} - i\frac{\partial}{\partial t}\right)\widetilde{e}_{\mathrm{a}}(t) + V^{*}\widetilde{\phi}(x_a, t) = 0 \tag{15}$$

$$\left(\omega_{\mathrm{c,b}} - i\frac{1}{\tau_{\mathrm{c,b}}} - i\frac{\partial}{\partial t}\right)\widetilde{e}_{\mathrm{b}}(t) + V^{*}\widetilde{\phi}(x_b, t) = 0 \tag{16}$$

The numerical results and a schematic of the system considered are shown in Fig. 7. We see that the conversion efficiency is 100% with a zero probability of detecting the original photon energy after the conversion process took place. This established that the wavelength conversion of the single photon is complete and that it leaves the photon wavepacket in tact.

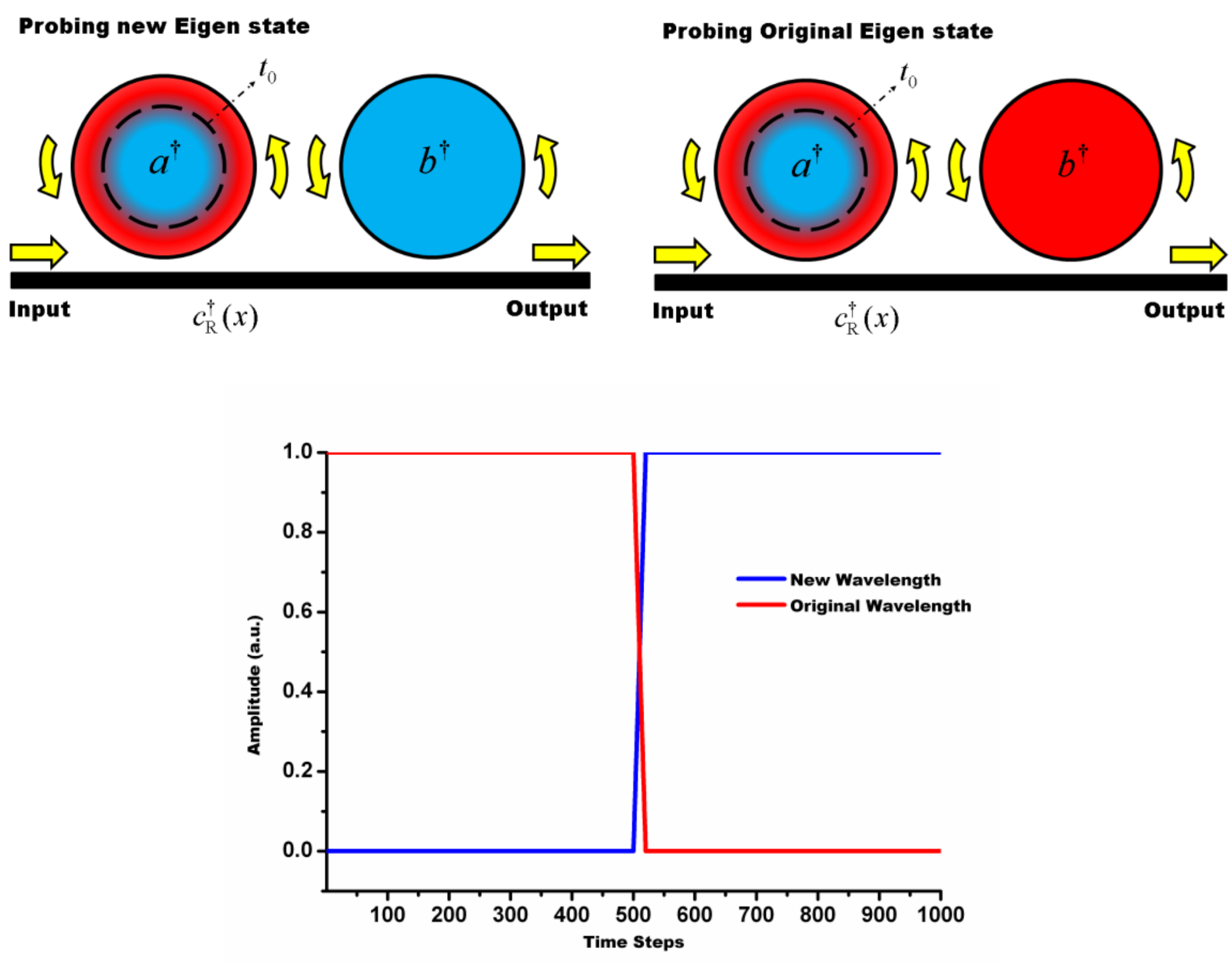

FIG. 7. The state inside the cavity changes with 100% efficiency final state of the cavity tuning. During the transition period the photon follows the state of the cavity.

## IV. SINGLE-PHOTON PACKET TRAPPING IN COHERENT POPULATION TRAPPING ANALOGUE SYSTEMS

### *a. Theoretical Model*

In this section we apply our dynamical analysis to study the coherent trapping of a single photon wave packet in a multi-waveguide multi-ring system. Similar classical light experiments have been performed which rely on a photonic EIT analogue [16,17,18,27,28]. Due to the importance of the quality of the storage unit we consider the structure in [16] where the interaction between the photon and the dynamic parts of the system is minimal. This opens the door for efficient single photon processing unit based on silicon electronic devices, enabling the integration of hybrid structures capable of a wider range of functions. A schematic of the device is shown in Fig. 8.

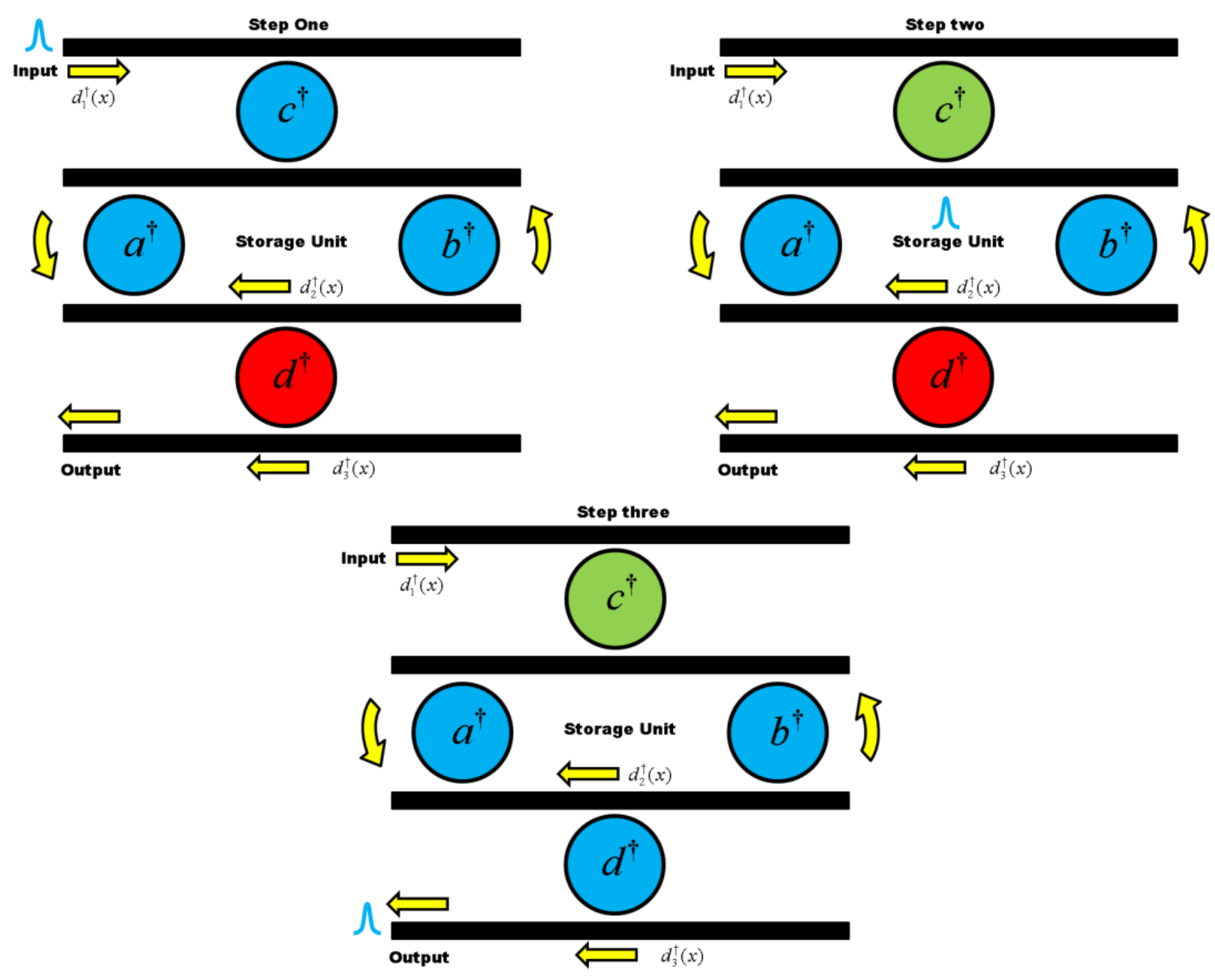

FIG. 8. Schematic the single photon storage unit, (Step 1) shows the acceptance state of the system. Bits are stored as shown in (Steps 2) then released in (Steps 3).

Equations (15)-(22) below are the time evolution equations for the system. Note that that we are considering microcavities with no internal coupling between clockwise and counter-clockwise modes so that there will be a uni-directional excitation of waveguide modes. This implies in Fig. 8 that modes $d_1$,$d_2$, $d_3$ ,and , $d_4$ propagate in the left, right, left, and right directions, respectively. Furthermore, the initial state of the system (input state position and direction) determine the permitted coupling conditions.

$$\left(\omega_0 - iv_g \frac{\partial}{\partial x_1} - i\frac{\partial}{\partial t}\right)\widetilde{\phi}_{d1}(x,t) + \delta(x_c)V\widetilde{e}_c(t) = 0 \tag{15}$$

$$\left(\omega_0 - iv_g \frac{\partial}{\partial x_2} - i\frac{\partial}{\partial t}\right)\widetilde{\phi}_{d2}(x,t) + \delta(x_a)V\widetilde{e}_a(t) + \delta(x_b)V\widetilde{e}_b(t) + \delta(x_c)V\widetilde{e}_c(t) = 0 \tag{16}$$

$$\left(\omega_0 - iv_g \frac{\partial}{\partial x_3} - i\frac{\partial}{\partial t}\right)\widetilde{\phi}_{d3}(x,t) + \delta(x_a)V\widetilde{e}_a(t) + \delta(x_b)V\widetilde{e}_b(t) = 0 \tag{17}$$

$$\left(\omega_0 - iv_g \frac{\partial}{\partial x_4} - i\frac{\partial}{\partial t}\right)\widetilde{\phi}_{d4}(x,t) + \delta(x_d)V\widetilde{e}_d(t) = 0 \tag{18}$$

$$\left(\omega_{\mathrm{c,a}} - i\frac{1}{\tau_{\mathrm{c,a}}} - i\frac{\partial}{\partial t}\right)\widetilde{e}_{\mathrm{a}}(t) + V^*\widetilde{\phi}_{d2}(x_a,t) + V^*\widetilde{\phi}_{d3}(x_a,t) = 0 \tag{19}$$

$$\left(\omega_{\mathrm{c,b}} - i\frac{1}{\tau_{\mathrm{c,b}}} - i\frac{\partial}{\partial t}\right)\widetilde{e}_{\mathrm{b}}(t) + V^*\widetilde{\phi}_{d2}(x_b,t) + V^*\widetilde{\phi}_{d3}(x_b,t) = 0 \tag{20}$$

$$\left(\omega_{\mathrm{c,c}}(t) - i\frac{1}{\tau_{\mathrm{c,c}}} - i\frac{\partial}{\partial t}\right)\widetilde{e}_{\mathrm{c}}(t) + V^*\widetilde{\phi}_{d2}(x_c,t) + V^*\widetilde{\phi}_{d3}(x_c,t) = 0 \tag{21}$$

$$\left(\omega_{\mathrm{c,d}} - i\frac{1}{\tau_{\mathrm{c,d}}} - i\frac{\partial}{\partial t}\right)\widetilde{e}_{\mathrm{d}}(t) + V^*\widetilde{\phi}_{d4}(x_d,t) + V^*\widetilde{\phi}_{d3}(x_d,t) = 0 \tag{22}$$

The system operates as follows; initially the input cavity is tuned to the storage unit cavities to direct the single photon wave packet into the system. Next the eigenstate of the input cavity is changed to trap the photon in the storage unit. In order to release the photon from the system we tune the eigenstate of the output cavity to the storage unit.

### b. *Results*

The system of equations were solved numerically with a single photon Gaussian packet as our input state propagating from left to right in waveguide mode $d_1$. First the packet is coupled to the storage unit then the storage unit is closed by detuning the input cavity and we see that we can hold the wavepacket. At later times the input cavity is tuned back to the storage unit and the packet leaks back to the output port as shown in Fig. 9 . In this way we have demonstrated the coherent storage and release of a single photon via linear optical interactions in a network microcavity system.

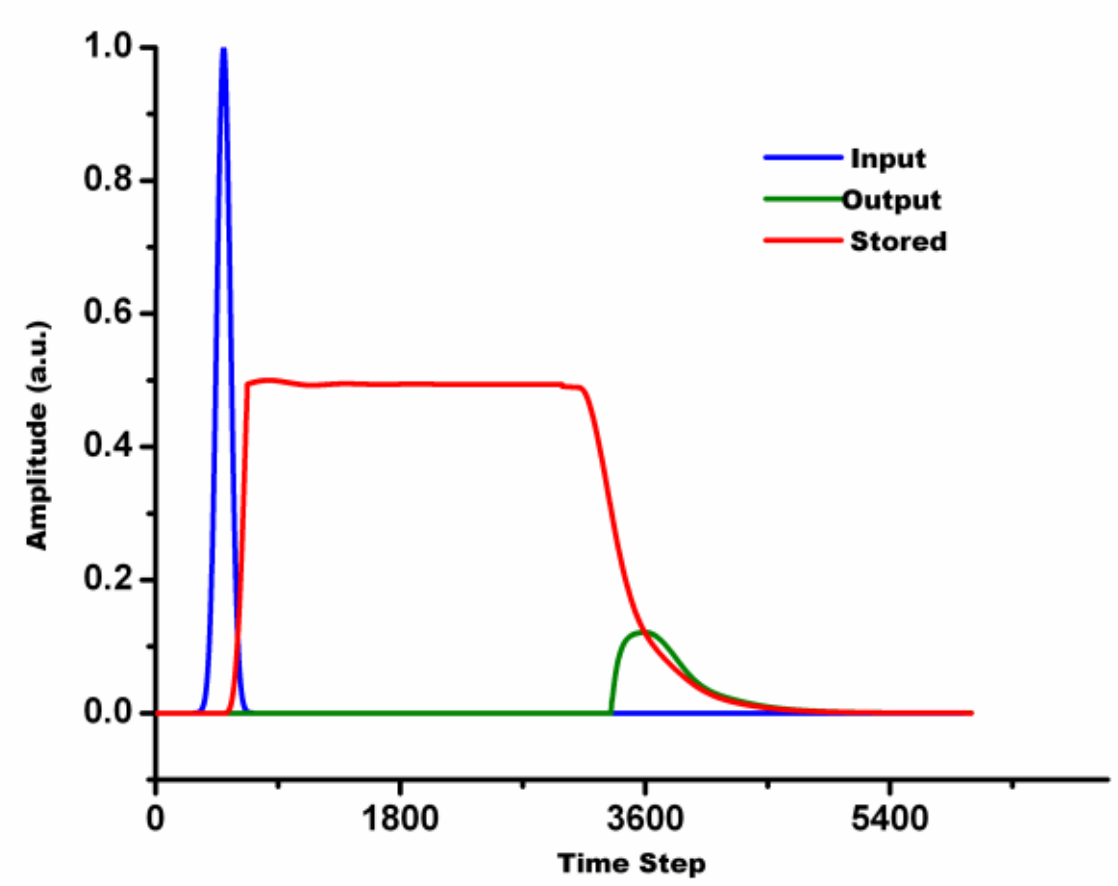

FIG. 9. Storage and release process of single photon wave-packet

## V. Outlook and Discussion

We have investigated the quantum dynamics of single photon transport through systems of waveguides coupled to microcavity resonators. Through this investigation we have established the single photon results for several systems of interest to advanced photonic systems. In all cases we have demonstrated a robust quantum mechanical foundation with analogous results to classical fields. Further we have presented the single photon case of a rigorous method for solving photonic transport problems in an "ansatz free" manner. We are now working to extend these analyses to the two and, eventually, *n*- photon cases. Finally, we have demonstrated theoretically the quantum mechanical behaviors of several photonic control mechanisms involving linear optical interactions. In addition, these results provide essential understanding of experimental demonstration of dynamic silicon photonic devices. We will continue to extend and broaden these studies both theoretically and experimentally through our future work.

**ACKNOWLEDGMENT**

The authors would like to thank Gernot Pomrenke, of the Air Force Office of Scientific Research, for his support through an AFOSR YIP award, and would like to acknowledge support from the National Science Foundation under award ECCS-0824103. We also thank Joseph Lobozzo II for the Lobozzo Optics Laboratory.

**APPENDIX**

Here we solve for the stationary states of the system without reference to any particular ansatz for the form of the result. We show that the well established results for the stationary states of the system emerge naturally from the mathematical structure of the approach. We anticipate that this method, properly extended, might simplify the significantly more complicated mathematical development of multi-photon transport processes involving quantum electrodynamic couplings between systems having continuous spectra and those having discrete spectra.

The general solution to the differential equation for $\phi(x)$ can be written as $\phi(x) = \phi_{\mathrm{h}}(x) + \phi_{\mathrm{p}}(x)$ where $\phi_{\mathrm{h}}(x) = \phi_{\mathrm{h}}(0)e^{iQx}$ is the homogeneous (viz. $V$ = 0) solution with $Q \equiv \frac{\omega-\omega_0}{v_g}$. We now obtain the particular solution, $\phi_{\mathrm{p}}(x)$.

Introducing the Fourier Transform, $F(\alpha) = \int_{-\infty}^{+\infty} dx\phi(x)e^{i\alpha x}$, and its inverse, $\phi(x) = \frac{1}{2\pi}\int_{-\infty}^{+\infty} d\alpha F(\alpha)e^{-i\alpha x}$, we transform Eqn. (6) into Fourier ($\alpha$) space to obtain

$$(\omega_0 - v_g\alpha - \omega)F(\alpha) + Ve_{\mathrm{cav}} = 0$$

where, owing to the trivial nature of the Fourier Transform of Eqn. (7), we obtain $e_{\mathrm{cav}} = \dfrac{V^*\phi(0)}{\omega - \omega_{\mathrm{c}} + i\frac{1}{\tau_{\mathrm{c}}}}$.

Combining these results we obtain for the Fourier Transform of the waveguide amplitude function,

$$F(\alpha)=\left(\frac{\frac{|V|^2}{v_g}\phi(0)}{\omega-\omega_{\mathrm{c}}+i\frac{1}{\tau_{\mathrm{c}}}}\right)\frac{1}{\alpha+\left[\frac{\omega-\omega_0}{v_g}\right]}$$

. Inverting the Fourier Transform yields a particular solution,

$$\phi_{\mathrm{p}}(x)=\frac{1}{2\pi}\left(\frac{\frac{|V|^2}{v_g}\phi(0)}{\omega-\omega_{\mathrm{c}}+i\frac{1}{\tau_{\mathrm{c}}}}\right)\int_{-\infty}^{+\infty}\frac{d\alpha e^{-i\alpha x}}{\alpha+\left(\frac{\omega-\omega_0}{v_g}\right)},$$

which we evaluate using the calculus of residues after analytically extending $F(\alpha)$ into the a complex $z$-plane, defined by $z\equiv\alpha+i\beta$ where $\alpha$ and $\beta$ are each real numbers. Let us work out the details by considering for a moment only the integral and defining, suggestively, $Q\equiv\left(\frac{\omega-\omega_0}{v_g}\right)$. That is, we must ascertain the value of the improper integral, $\int_{-\infty}^{+\infty}\frac{d\alpha e^{-i\alpha x}}{\alpha+Q}$, where $q$ is a real number. Clearly the integrand has a simple pole at $\alpha=-Q$.

In order to apply the residue theorem from complex analysis, we define the contour integral around a closed contour in the z-plane, $\oint\frac{dze^{-izx}}{z+Q}=P\int_{-\infty}^{+\infty}\frac{d\alpha e^{-i\alpha x}}{\alpha+Q}+\lim_{\eta\to0}\int_{\mathrm{sc}}\frac{dze^{-izx}}{z+Q}+\int_{\mathrm{C}}\frac{dze^{-izx}}{z+Q}$ , where $P$ indicates the Cauchy Principal Value of the improper integral and C is an infinite semicircle chosen to ensure convergence and therefore vanishing of this contribution to the contour integral. The remaining term represents an infinitesimal semicircular deformation in the contour needed to accommodate a pole such as this one that lies along the path of the integration [29]. The various pieces of the contours for the cases $x > 0$ and $x < 0$ are displayed in Figure (10). Referring to Figure (10) and using the result that the residue theorem as applied here gives $\oint\frac{dze^{-izx}}{z+Q}=\pm 2\pi ie^{iQx}$ provided that the pole at $\alpha = -Q$ is encircled by the contour of integration and where the + (-) sign corresponds to a counterclockwise (clockwise) sense of integration. Applying the mean value theorem to the infinitesimal semicircular deformation yields $\lim_{\eta \to 0} \int_{\text{sc}} \frac{dz e^{-izx}}{z+Q} = \pm \pi i e^{iQx}$ where the + (-) sign corresponds to a counterclockwise (clockwise) sense of integration around the infinitesimal *deformation* in the contour.

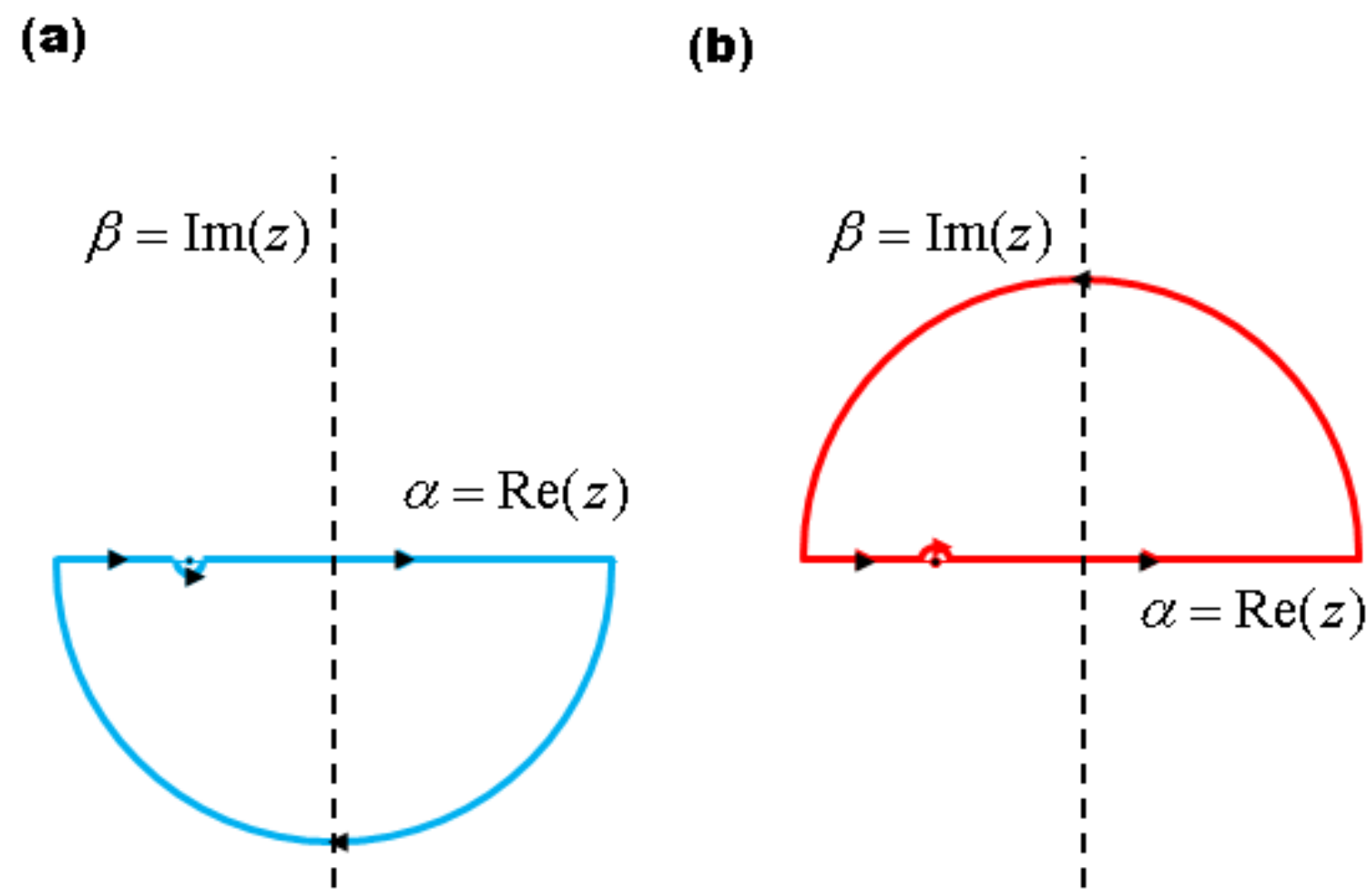

FIG. 10. Contours in the complex z plane used for inversion of the Fourier transform of the waveguide amplitude for the cases (a) $x > 0$ and (b) $x < 0$.

We now handle the integration in two regions. For $x > 0$, we choose the clockwise path having the infinite semicircular branch in the $\beta < 0$ half plane (see Fig. (10a)). Deforming the contour in a counterclockwise sense relative to the pole at $\alpha$ = -$Q$ *excludes* the pole from the region encircled by the contour. So, for $x > 0$, we obtain $P\int_{-\infty}^{+\infty} \frac{d\alpha e^{-i\alpha x}}{\alpha+Q} = -i\pi e^{iQx}$ (note that including the pole with a clockwise deformation yields the same result). For $x < 0$, we choose the counterclockwise path having the infinite semicircular branch in the $\beta > 0$ half plane (see Fig. (10b)). Deforming the contour in a clockwise sense relative to the pole at $\alpha$ = -$Q$ *excludes* the pole from the region encircled by the contour. So, for $x < 0$, we obtain $P\int_{-\infty}^{+\infty} \frac{d\alpha e^{-i\alpha x}}{\alpha+Q} = i\pi e^{iQx}$ (again independent of the choice for the deformation).

Collecting the results from the calculus of residues and combining them in the inverse Fourier form for the particular solution gives $\phi_{\mathrm{p}}(x)=\frac{i}{2}e^{iQx}\left(\frac{\frac{|V|^2}{v_g}\phi(0)}{\omega-\omega_{\mathrm{c}}+i\frac{1}{\tau_{\mathrm{c}}}}\right)[\theta(-x)-\theta(x)]$. Notice that the residue theorem immediately enforces the appropriate dispersion relation for our *rightward* traveling wave. Now, it is clear from the forms of $\phi_{\mathrm{h}}(x)$ and $\phi_{\mathrm{p}}(x)$ that $\phi_{\mathrm{p}}(0)=0$ and $\phi_{\mathrm{h}}(0)=\phi(0)$. Defining $\phi(0)\equiv\phi_0$, $\Gamma\equiv\frac{|V|^2}{2v_g}$, and $D\equiv\omega-\omega_{\mathrm{c}}+i\frac{1}{\tau_{\mathrm{c}}}$ to simplify the notation, we can now write the general solution for the stationary state as $\phi(x)=\phi_0e^{iQx}\left(1+\frac{i\Gamma}{D}[\theta(-x)-\theta(x)]\right)=\phi_0e^{iQx}\left[\left(1+\frac{i\Gamma}{D}\right)\theta(-x)+\left(1-\frac{i\Gamma}{D}\right)\theta(x)\right]$. This state has the form $\phi(x)=e^{iQx}[A\theta(-x)+B\theta(x)]=\begin{cases}Ae^{iQx} & x<0\\ Be^{iQx} & x>0\end{cases}$, exactly the form of a one dimensional scattering problem in which (i) the target is localized at $x$ = 0, (ii) there is no reflection ($r$ = 0), and (iii) the target has internal structure from which irreversible losses to the environment can occur. In one dimensional scattering theory the transmission coefficient is defined as $t\equiv\frac{B}{A}$ as it relates to this form. In the present case we have $t\equiv\frac{D-i\Gamma}{D+i\Gamma}$ which, upon substitution for $D$ and $\Gamma$, is exactly the result given in Eqn. (9) in Section II.

We emphasize that the method we have presented in this Appendix relies on no additional assumption beyond the existence of the Fourier transform pairs we have used. Rather, the form of the interacting eigenstate emerges naturally as the solution to the system of equations determining the stationary states. In fact, the state that we derive as the general solution turns out to be exactly the single particle Bethe ansatz state that one would expect in this simple case. In a future work we will extend this method to cases involving more than one photon, and in so doing provide another mathematical mechanism for backing out the S-matrix for the interaction. In the single photon case and; therefore, for our purposes here, this further sophistication is not necessary.

**REFFERENCES**

[1] J.L. O'Brien, A. Furusawa, and J. Vuckovic, "Photonic quantum technologies," *Nature Photonics*, vol. 3, 2009, pp. 687 - 695.

[2] S. Clemmen, K. Phan Huy, W. Bogaerts, R.G. Baets, P. Emplit, and S. Massar, "Continuous wave photon pair generation in silicon-on-insulator waveguides and ring resonators.," *Optics express*, vol. 17, 2009, pp. 16558-70.

[3] M.A. Nielsen and I.L. Chuang, *Quantum Computation and Quantum Information*, Cambridge University Press, 2000.

[4] W. Tittel, J. Brendel, H. Zbinden, and N. Gisin, "Quantum cryptography using entangled photons in energy-time bell states," *Physical review letters*, vol. 84, 2000, pp. 4737-40.

[5] T. Nagata, R. Okamoto, J.L. O'brien, K. Sasaki, and S. Takeuchi, "Beating the standard quantum limit with four-entangled photons.," *Science (New York, N.Y.)*, vol. 316, 2007, pp. 726-9.

[6] B.L. Higgins, D.W. Berry, S.D. Bartlett, H.M. Wiseman, and G.J. Pryde, "Entanglement-free Heisenberg-limited phase estimation.," *Nature*, vol. 450, 2007, pp. 393-6.

[7] Y. Kawabe, H. Fujiwara, R. Okamoto, K. Sasaki, and S. Takeuchi, "Quantum interference fringes beating the diffraction limit.," *Optics express*, vol. 15, 2007, pp. 14244-50.

[8] a. Boto, P. Kok, D. Abrams, S. Braunstein, C. Williams, and J. Dowling, "Quantum interferometric optical lithography: exploiting entanglement to beat the diffraction limit," *Physical review letters*, vol. 85, 2000, pp. 2733-6.

[9] J.L. O'Brien, "Optical quantum computing.," *Science (New York, N.Y.)*, vol. 318, 2007, pp. 1567-70.

[10] E. Knill, R. Laflamme, and G.J. Milburn, "A scheme for efficient quantum computation with linear optics.," *Nature*, vol. 409, 2001, pp. 46-52.

[11] J. Shen and S. Fan, "Theory of single-photon transport in a single-mode waveguide. I. Coupling to a cavity containing a two-level atom," *Physical Review A*, vol. 79, 2009, pp. 1-11.

[12] J. Shen and S. Fan, "Theory of single-photon transport in a single-mode waveguide. II. Coupling to a whispering-gallery resonator containing a two-level atom," *Physical Review A*, vol. 79, 2009.

[13] M. Notomi, T. Tanabe, E. Kuramochi, A. Shinya, and H. Taniyama, "Photonic Crystal Nanocavities : Slow Light , All-optical Processing , Wavelength Conversion , Optical MEMS," *Group IV Photonics*, 2007.

[14] S.F. Preble, Q. Xu, and M. Lipson, "Changing the colour of light in a silicon resonator," *Nature Photonics*, vol. 1, 2007, pp. 293-296.

[15] Z. Gaburro, M. Ghulinyan, F. Riboli, L. Pavesi, A. Recati, and I. Carusotto, "Photon energy lifter," *Opt. Express*, vol. 14, 2006, pp. 7270-7278.

[16] A.W. Elshaari, A. Aboketaf, and S.F. Preble, "Controlled storage of light in silicon cavities.," *Optics express*, vol. 18, 2010, pp. 3014-22.

[17] Q.F. Xu, P. Dong, and M. Lipson, "Breaking the delay-bandwidth limit in a photonic structure," *Nature Phys.*, vol. 3, 2007, pp. 406-410.

[18] M.F. Yanik and S. Fan, "Slow light: Dynamic photon storage," *Nature Phys.*, vol. 3, 2007, pp. 3-5.

[19] Y. Takahashi, H. Hagino, Y. Tanaka, B.S. Song, T. Asano, and S. Noda, "High-Q nanocavity with a 2-ns photon lifetime," *Optics Express*, vol. 15, 2007, pp. 17206-17213.

[20] J. Sakurai, *Modern Quantum Mechanics*, Addison Wesley Longman, 1994.

[21] J. Shen and S. Fan, "Strongly correlated multiparticle transport in one dimension through a quantum impurity," *Physical Review A*, vol. 76, 2007.

[22] C.R. Otey, M.L. Povinelli, and S. Fan, "Capturing light pulses into a pair of coupled photonic crystal cavities," *Applied Physics Letters*, vol. 94, 2009, p. 231109.

[23] A. Taflove and S. Hagness, *Computational Electrodynamics:*, Artech House Publishers, 2005.

[24] M. Notomi and S. Mitsugi, "Wavelength conversion via dynamic refractive index tuning of a cavity," *Phys. Rev. A*, vol. 73, 2006, p. 051803.

[25] David J. Griffith, *Introduction to Quantum Mechanics*, Prentice Hall, .

[26] P. Dong, S. Preble, J. Robinson, S. Manipatruni, and M. Lipson, "Inducing Photonic Transitions between Discrete Modes in a Silicon Optical Microcavity," *Physical Review Letters*, vol. 100, 2008, pp. 1-4.

[27] Q. Xu, S. Sandhu, M. Povinelli, J. Shakya, S. Fan, and M. Lipson, "Experimental Realization of an On-Chip All-Optical Analogue to Electromagnetically Induced Transparency," *Physical Review Letters*, vol. 96, 2006, pp. 1-4.

[28] J. Upham, Y. Tanaka, T. Asano, and S. Noda, "Dynamic increase and decrease of photonic crystal nanocavity Q factors for optical pulse control.," *Optics express*, vol. 16, 2008, pp. 21721-30.

[29] J.H. Mathews, *Complex Variables for Mathematics and Engineering*, Wm. C. Brown Publishers, 1988.